\documentclass{article}
\usepackage[left=1in, right=1in, top=1in, bottom=1in]{geometry}
\usepackage{amsmath,amsfonts}
\usepackage{algorithmic}
\usepackage{algorithm}
\usepackage{enumitem}
\usepackage{array}
\usepackage{textcomp}
\usepackage{stfloats}
\usepackage{url}
\usepackage{verbatim}
\usepackage{graphicx}
\usepackage{caption}
\usepackage{subcaption}
\usepackage{cite}
\usepackage{soul,color}
\usepackage{multirow} % Include this in your preamble
\usepackage{amsmath}
\usepackage{adjustbox}
\usepackage{array}
\usepackage{flushend} % Added for leveling the references

\usepackage{authblk}

\title{Lightweight CNN-Based Wi-Fi Intrusion Detection Using 2D Traffic Representations}
\author{Rayed Suhail Ahmad}
\author{Rehan Ahmad}
\author{Quamar Niyaz\thanks{Corresponding author: Quamar Niyaz (qniyaz@pnw.edu)}}
\affil{Electrical and Computer Engineering Department, Purdue University Northwest, Hammond, IN 46323}
\date{}

\begin{document}

\maketitle
\begin{abstract}
Wi-Fi networks are ubiquitous in both home and enterprise environments, serving as a primary medium for Internet access and forming the backbone of modern IoT ecosystems. However, their inherent vulnerabilities, combined with widespread adoption, create opportunities for malicious actors to gain unauthorized access or compromise sensitive data stored on connected devices. To address these challenges, we propose a deep learning based network intrusion detection system (NIDS) for Wi-Fi environments. Building on our previous work, we convert network traffic into two-dimensional data representations and use them to train DL models based on convolutional neural network (CNN) architectures. We implement five distinct techniques for generating the two-dimensional representations, and to ensure low detection latency, we adopt lightweight CNN architectures in our NIDS. The models are trained using the AWID3 dataset, a publicly available benchmark for Wi-Fi NIDS research, and are evaluated for both binary and multi-class classification tasks. Experimental results demonstrate that the proposed approach achieves competitive detection performance with low inference time, making it suitable for real-world Wi-Fi deployment scenarios.   
\end{abstract}


\section{Introduction}
\label{sec:intro}
The number of Internet users reached approximately 5.5 billion at the beginning of 2025, accounting for about 67\% of the global population \cite{statistainternet}. With the proliferation of portable devices (e.g., smartphones, laptops), Wi-Fi has become one of the primary technologies enabling Internet connectivity. Furthermore, the rapid expansion of IoT ecosystems—including home automation, healthcare monitoring, and industrial automation—has significantly increased Wi-Fi usage. For instance, 45\% of U.S. households with Internet access own at least one smart home device. The total number of active Wi-Fi devices worldwide reached 21.1 billion in 2024 \cite{idcwifi}. However, the convenience and flexibility that Wi-Fi provides to Internet users and the IoT ecosystem also introduce significant security challenges, creating opportunities for malicious actors to exploit vulnerabilities and launch targeted attacks.

Adversaries exploit weaknesses in Wi-Fi design and configuration, with public Wi-Fi networks being a prime target. Simple exploits can escalate into advanced Man-in-the-middle (MITM) attacks, such as the evil twin attacks \cite{ghering2016evil}, which are increasingly common and inexpensive to execute. Nearly 40\% of surveyed US adults reported data compromise from public Wi-Fi use \cite{public_wifi}. A recently disclosed vulnerability, SSID (Service Set Identifier) confusion, enables attackers to trick victims into connecting to insecure networks and intercept their traffic \cite{gollier}. IoT systems have seen a 124\% rise in attacks in 2024, including IP Camera Command Injection and the Reaper botnet attacks \cite{lacoma}. A massive breach exposed 2.7 billion records containing Wi-Fi SSIDs, passwords, IP addresses, and device IDs \cite{mascellino}, enabling unauthorized access and “nearest neighbor” exploits, where attackers can intrude nearby Wi-Fi networks. Such attacks threaten the confidentiality, integrity, and availability of critical data.

Various defense mechanisms are deployed to detect and mitigate attacks in both home and enterprise networks. Among these, the Network Intrusion Detection System (NIDS) plays a critical role by continuously monitoring network traffic for signs of malicious activity or anomalous behavior. Traditional NIDSs, such as Snort \cite{snort} and Suricata \cite{suricata}, rely on signature-based detection, which matches traffic against predefined attack patterns. While effective for known threats, this approach is inherently limited in identifying zero-day attacks. To address these limitations, recent research integrates state-of-the-art machine learning (ML) techniques into NIDSs. ML-based models can learn complex traffic patterns from data, enabling them to identify previously unseen attacks—including zero-day threats—without the need for predefined signatures. 

ML-based approaches have achieved significant success across various domains (e.g., computer vision, recommendation systems). Within ML, a prominent subset known as Deep Learning (DL) leverages neural networks to automatically extract high-level features from data—such as images and audio—without the need for handcrafted feature engineering \cite{lecun2015deep}. These techniques have demonstrated exceptional performance in tasks like object detection and image classification \cite{pouyanfar2018survey}. Motivated by these advancements, our work proposes an NIDS that processes Wi-Fi network traffic transformed into a two-dimensional representation analogous to images, enabling DL-based classification for intrusion detection in Wi-Fi networks. The primary motivation for this approach is to assess the effectiveness of combining two-dimensional representations of Wi-Fi traffic with DL-based classification for building an NIDS capable of identifying malicious activities in the network. We investigate multiple transformation techniques alongside lightweight convolutional neural network (CNN) architectures—a widely adopted DL method—for our NIDS. Model training is performed on a subset of the Aegean Wi-Fi Intrusion Dataset 3 (AWID3) \cite{awid3_2021}, a publicly available dataset containing diverse samples of normal and anomalous Wi-Fi network traffic. The anomalous traffic in the AWID3 dataset was generated by executing a variety of attacks on an enterprise-grade Wi-Fi network. We compare both detection performance and inference time across our different techniques to design a robust and responsive NIDS capable of identifying malicious intrusions in Wi-Fi networks. This comparative evaluation enables us to systematically analyze the trade-offs between detection accuracy and speed, ultimately selecting an optimal NIDS configuration for deployment. 

This work builds upon and substantially extends our previous study \cite{rayed}, which employed Gramian Angular Field (GAF) transformation of Wi-Fi traffic in conjunction with a CNN architecture to develop a Wi-Fi NIDS. In this extension, we design models capable of classifying Wi-Fi traffic into attack and normal categories (binary classification) as well as categorizing attacks into their respective types (multi-class classification). We apply five distinct transformation techniques to convert Wi-Fi traffic features into two-dimensional representations, enabling a comprehensive performance evaluation. For the 1D-CNN architecture, the input format is modified by flattening the two-dimensional representations into a one-dimensional form. Furthermore, the literature review has been substantially revised to focus exclusively on related works utilizing the AWID3 dataset, ensuring a more precise and relevant comparison.  

Toward this end, the remainder of the paper is organized as follows. Section \ref{sec:relatedwork} reviews prior works that have implemented Wi-Fi NIDS using the AWID3 dataset, along with selected studies employing two-dimensional transformation techniques for network traffic representation. Section \ref{sec:implementation} presents the implementation details of the proposed Wi-Fi NIDS, including the transformation techniques and CNN architectures, as well as a brief description of the AWID3 dataset and the specific features utilized in this work. Section \ref{sec:results} reports the experimental results based on the defined performance metrics. Finally, Section \ref{sec:conclusion} provides concluding remarks.

\section{Related Work}
\label{sec:relatedwork}
In this section, we will briefly review the recent literature that utilized the AWID3 dataset to develop their ML based Wi-Fi IDS. In addition, we will discuss a few works that utilize imaging or two dimensional techniques for performing similar tasks. In \cite{multiple_ml_awid3_2021}, Anibal assessed the performance of various online learning classifiers, including Hoeffding Tree (HT), K-nearest Neighbors (KNN), Accuracy Weighted Ensemble Classifier (AWEC), and Half-Spaced Treed (HST), for detecting attacks in WiFi networks. The study utilized both the AWID2 and AWID3 datasets, implementing models to classify network traffic into 4 and 2 distinct classes, respectively. Experimental results demonstrated that HT achieved the highest accuracy of 98\% for binary classification for AWID3 dataset using a feature set of 10 attributes. 

Chatzoglou et al. in \cite{awid_features_2022} employed 16 features extracted from Wi-Fi headers to train both shallow ML and DL models for multi-class classification into \textit{Normal, Flooding,} and \textit{Impersonation} categories. Among the shallow ML models, the Extra Trees (ET) classifier achieved the best performance, with an accuracy of 99.96\% and an F1-score of 99.52\% on the AWID3 dataset. For DL-based approaches, a Multi-layer Perceptron (MLP) and an Auto-Encoder (AE) were used, yielding accuracies of 99.73\% and 99.66\%, and F1-scores of 97.55\% and 96.78\%, respectively, on the same dataset. The authors in \cite{ensemble_ml_awid3_2022} utilized the same set of 16 features derived from 802.11 Wi-Fi headers as proposed in \cite{awid_features_2022}, along with an additional 17 non-802.11 features extracted from ARP, IP, TCP, and UDP protocol headers, to detect application-layer attacks (\textit{Botnet, Malware, SSH Brute Force, SQL Injection, SSDP Amplification,} and \textit{Website Spoofing}). They trained both shallow and DL models to classify network traffic into three categories: \textit{Normal, Flooding} (comprising SSDP Amplification, SSH Brute Force, and Website Spoofing), and \textit{Other}. The bagging classifier among the shallow models demonstrated the best performance, achieving an accuracy of 98.73\% and an F1-score of 88.07\% when using only the 802.11 feature set.

In \cite{bhutta2024lightweight}, the authors proposed a lightweight Wi-Fi IDS based on Light Gradient Boosting Machine (LightGBM) algorithm, which combines multiple weak learners to form a strong predictive model. To enhance training efficiency, they employed Gradient-based One-Side Sampling to down-sample data instances associated with small gradients. Using 17 features extracted from the AWID3 dataset, their model achieved an accuracy of 99.77\% and an F1-score of 94.79\%. A multi-class classification approach was implemented to distinguish between 19 Wi-Fi traffic classes, comprising 18 attack types and normal traffic. The proposed method demonstrated superior performance, achieving comparable accuracy with 26 times faster training and 20\% lower test time compared to the XGBoost algorithm. Bhutta et al. in \cite{bhutta2025advancing} proposed a custom oversampling method, Balanced Boost Oversampling Technique (BBOT), to address class imbalance, generate synthetic samples, and enhance model performance. Experimental results demonstrated that the integration of feature selection, BBOT, and a Decision Tree classifier outperformed all other models while maintaining low computational overhead, offering an efficient solution for intrusion detection in WiFi networks. Their approach achieved comparable performance to XGBoost, but with a significantly reduced training time of just 56 seconds. They reported an accuracy of 99.45\% and an F1-score of 97\% in classifying 19 traffic classes, consistent with the previously discussed work.

In \cite{kampourakis2025balancing}, Kampourakis et al. investigated the effect of dataset balance on the performance of a wireless IDS developed using various ML algorithms. Their study utilized the AWID3 dataset in a multi-class IDS setting, targeting three categories: normal traffic, flooding attacks, and impersonation attacks. A total of 10 ML algorithms were evaluated, comprising both traditional (shallow) models and DNNs. Given the inherent imbalance in AWID3, three resampling strategies were applied: undersampling, oversampling, and a hybrid approach termed “overundersampling”. The highest performance was observed using the imbalanced dataset with the shallow LightGBM model, leveraging 16 traffic features. This configuration achieved an accuracy of 99.96\% and an F1-score of 98.95\%. Yonbawi et al. \cite{yonbawi2025transferability} investigated the transferability of IDS ML models to novel and previously unseen datasets. To this end, they trained Convolutional Neural Network (CNN) and Multilayer Perceptron (MLP) models on the AWID2 dataset and evaluated their performance on the AWID3 dataset. The CNN-based model demonstrated superior performance compared to the MLP model. During training on the AWID2 dataset, the CNN achieved an F1-score of 99.96 and an accuracy of 99.94\% using 8 selected features. When evaluated on the AWID3 dataset, the pre-trained CNN model achieved an accuracy of 97.28\% and an F1-score of 98.53\% in a binary classification task distinguishing between normal traffic and flooding attacks.

In \cite{seq2img_2017}, Chen et al. proposed a method for classifying application types from IP traffic flows by transforming IP packets into images and processing them through a CNN-based model. From the bidirectional flow of IP packets, features such as packet size sequences, inter-arrival times, and directions were embedded as images using the Reproducing Kernel Hilbert Space (RKHS) mapping. The CNN architecture comprised two convolution–pooling layer pairs, followed by two fully connected hidden layers, and concluded with an output layer that classified the IP traffic into one of five application categories. Aminanto et al. \cite{jpeg_typing_2022} introduced a projection-based imaging approach that transforms network traffic into image form and classifies it using the CNN architecture EfficientNet \cite{efficient_net_2020}. Their work utilized the AWID2 dataset \cite{awid2_2015}, the predecessor of AWID3, to implement an IDS capable of distinguishing normal traffic from impersonation, injection, and flooding attacks. The method generated RGB (3-channel) images with a resolution of $224 \times 224$, structured as an $N \times N$ grid, where $N$ is the square root of $k$ selected features from each traffic record. Each grid cell was populated in a zig-zag order with a feature value rendered using the Hersley Simplex font. Experimental results indicated that EfficientNet achieved its best performance with $49$ features, yielding an F1 score of $99.94\%$. When compared to other models (RF, SVM, and XGBoost), the average performance rankings across different values of $k$ were RF, XGBoost, EfficientNet, and SVM. Notably, the performance gap among the top three models was less than $0.5\%$, indicating that the proposed approach delivered performance comparable to the strongest competing models. In a related study \cite{red_blue_img_2022}, Aminanto et al. applied feature extraction using Recursive Feature Elimination and Cross-Validation (RFECV) and transformed the AWID2 tabular dataset into RGB image representations, utilizing only the red and blue channels—an approach inspired by \cite{cnn_unsw_img_2021}—to develop a deep learning-based IDS. Model training was conducted using a ResNet50 \cite{resnet_2015} base model. Their highest performance was obtained with $64$ features selected via RFECV, achieving an F1-score of $99.73\%$. Furthermore, when comparing the best-performing CNN model against alternative classifiers such as SVM, DT, and XGBoost, the results indicated that the CNN achieved performance comparable with state-of-the-art models.

In this work, we build upon the findings of Chatzoglou et al. \cite{awid_features_2022}, which provide a set of significant features extracted from Wi-Fi frames for the NIDS. Unlike their approach, which used the AWID2 dataset in its original imbalanced form, directly applying the same method to AWID3 would be computationally expensive due to its size of roughly $43$ GB, demanding substantial memory and processing time for model training. To address this, we sample the dataset to mitigate the imbalance between normal and attack traffic. Although oversampling—shown to improve NIDS performance in \cite{imbalanced_1dcnn_2020} by Azizjon et al.—was considered, it was deemed impractical given AWID3’s severe class imbalance, where normal-to-attack traffic records occur in a $38:1$ ratio. Instead, we apply random undersampling, as suggested in \cite{over_under_sampling_2006}, reducing the ratio to $8:1$ for a more balanced and computationally manageable dataset. 

\section{Implementation of Wi-Fi NIDS}
\label{sec:implementation}
In the proposed Wi-Fi NIDS, network traffic is continuously captured from the Wi-Fi network. Relevant frame headers from the Wi-Fi link layer are then extracted, preprocessed, transformed, and subsequently fed into the deep learning-based model for intrusion detection. In this work, we assume that the Wi-Fi traffic has already been captured and that the corresponding frame headers have been extracted. Accordingly, this section focuses on the detailed implementation of the deep learning model for our IDS.
\subsection{Dataset Preparation and Preprocessing}
To implement the IDS, we used the AWID3 dataset \cite{awid3_2021}, a publicly available network intrusion dataset that consists of a broad range of recent Wi-Fi and multi-layered attacks, such as KRACK and Kr00k that were not present in its predecessor dataset, AWID2 \cite{awid2_2015}. The AWID3 dataset was collected in a testbed emulating a typical enterprise Wi-Fi network environment. The dataset offers a rich collection of labeled samples for both benign and malicious traffic, encompassing a wide variety of attack types. 
\begin{table}
\centering
\footnotesize
\caption{Distribution of traffic records in AWID3 dataset \cite{awid3_2021}}
\label{tab:awid3_division}
%\resizebox{\columnwidth}{!}{
\begin{tabular}{llll}
    \hline
    \textbf{Attack Type} & \textbf{Normal} & \textbf{Attack} & \textbf{Total} \\ \hline \hline
    De-Authentication & 1,587,527 & 38,942 & 1,626,469
    \\ \hline
    Disassociation & 1,938,585 & 7,531 & 1,946,116
    \\ \hline
    Re-Association & 1,838,430 & 5,502 & 1,843,932
    \\ \hline
    Rogue AP & 1,971,875 & 1,310 & 1,973,185
    \\ \hline
    Krack & 1,388,498 & 49,990 & 1,438,488
    \\ \hline
    Kr00k & 2,708,637 & 191,803 & 2,900,440
    \\ \hline
    Evil Twin & 3,673,854 & 104,827 & 3,778,681
    \\ \hline \hline
    SSH Brute Force & 2,428,688 & 11,882 & 2,440,570
    \\ \hline
    Botnet & 3,169,167 & 56,891 & 3,226,058
    \\ \hline
    Malware & 2,181,148 & 131,611 & 2,312,759
    \\ \hline
    SQL Injection & 2,595,727 & 2,629 & 2,598,356
    \\ \hline
    SSDP Amplification & 2,641,517 & 5,499,851 & 8,141,368
    \\ \hline
    Website Spoofing & 2,263,446 & 405,121 & 2,668,567
    \\ \hline
\end{tabular}
%}
\end{table}

Table \ref{tab:awid3_division} presents the distribution of normal and attack traffic in the AWID3 dataset. The dataset is provided in two formats: raw PCAP files and pre-processed CSV files. Each PCAP file contains Wi-Fi frame data captured corresponding to a specific attack or normal traffic scenario, which can be analyzed using network analysis tools such as Wireshark or Tcpdump. The accompanying CSV files consist of 253 features extracted from the PCAP data, along with an additional field that labels each traffic record as either normal or a specific attack type. In this work, we focus exclusively on the first seven attack types listed in Table \ref{tab:awid3_division}, as these are directly related to link-layer vulnerabilities in Wi-Fi networks.
\begin{table}
\centering
\footnotesize
\caption{Selected features with pre-processing steps \cite{rayed}}
\label{tab:Data_Processing}
%\resizebox{\columnwidth}{!}{
\begin{tabular}{llp{4cm}}
    \hline
    \textbf{Feature \#} & \textbf{Feature} & \textbf{Pre-processing step} \\ \hline \hline
    0 & frame.len & Min-Max Scaling \\ \hline
    1 & radiotap.length & Min-Max Scaling \\ \hline
    2 & radiotap.dbm\_antsignal & Max DBM value followed by Min-Max Scaling
    \\ \hline
    3 & wlan.duration & Min-Max Scaling \\ \hline
    4 & radiotap.present.tsft & One-Hot Encoding \\ \hline
    5 & radiotap.channel.freq & Min-Max Scaling  \\ \hline
    6 & radiotap.channel.flags.cck & Min-Max Scaling \\ \hline
    7 & radiotap.channel.flags.ofdm & Min-Max Scaling \\ \hline
    8 & wlan.fc.type & Min-Max Scaling \\ \hline
    9 & wlan.fc.subtype & Min-Max Scaling \\ \hline
    10 & wlan.fc.ds & Hexadecimal to Decimal conversion followed by Min-Max Scaling \\ \hline
    11 & wlan.fc.frag & Min-Max Scaling \\ \hline
    12 & wlan.fc.retry & Min-Max Scaling \\ \hline
    13 & wlan.fc.pwrmgt & Min-Max Scaling \\ \hline
    14 & wlan.fc.moredata & Min-Max Scaling \\ \hline
    15 & wlan.fc.protected & Min-Max Scaling \\ \hline
    16 & attack\_map & One-Hot Encoding \\ \hline
\end{tabular}
%}
\end{table}

As discussed, the original AWID3 dataset is structured into multiple CSV files based on traffic type—normal or specific attack categories. As a first step in data preparation, these files were consolidated into a single CSV file to form a unified dataset containing both normal and various attack traffic records. As evident from Table \ref{tab:awid3_division}, there exists a substantial imbalance between normal and attack traffic, with a ratio of approximately 38:1. To enhance computational efficiency during model training and address this class imbalance, we randomly sampled the normal traffic to reduce the ratio to approximately 8:1. Subsequently, traffic labels were encoded as numeric values: for binary classification, normal and attack traffic were assigned class labels 0 and 1, respectively. For multi-class classification, the label mapping was defined as follows: \textit{Normal} (0), \textit{De-Authentication} (1), \textit{Disassociation} (2), \textit{Re-Association} (3), \textit{Rogue AP} (4), \textit{Krack} (5), \textit{Kr00k} (6), \textit{Evil Twin} (7). 

Additionally, not all features contribute meaningfully to the detection of anomalous traffic, making feature selection a crucial step. According to findings in \cite{awid_features_2022}, models trained on a carefully selected subset of 16 features—out of the original 253—is able to achieve a high detection performance. In this work, we adopted the same 16 features in an effort to reduce computational complexity and enhance the efficiency of the Wi-Fi intrusion detection system with appropriate transformations applied to each feature. Table \ref{tab:Data_Processing} enumerates these 16 features and outlines the preprocessing steps applied to each feature. Specifically, features subjected to Min-Max Scaling were normalized within the range $\left[-1, 1\right]$ to prevent issues such as exploding gradients during training. 

The dataset was then partitioned into features and labels, followed by a stratified split into training, validation, and test sets to preserve class distribution. Initially, 70\% of the data was allocated to training and 30\% to testing. The training set was further split into 70\% training and 30\% validation subsets. The resulting datasets—training, validation, and test—were saved as separate CSV files for model development and evaluation.
\subsection{Two-dimensional Transformation Techniques}
Following data preparation and preprocessing, various two-dimensional transformation techniques were applied within the ML pipeline to train CNN models for both binary and multi-class classification tasks. Each technique converts a traffic record into a 2D matrix represented as grayscale pixel values, producing a single-channel image corresponding to that record. The initial step in each transformation technique involves mapping the feature indices from the dataset to specific pixel locations in the 2D matrix, followed by applying the computations specific to the given technique if needed. As stated earlier, Table \ref{tab:Data_Processing} provides us with the feature index assignments for the 16 selected features used in this study. The detailed discussion of the 2D transformation techniques utilized in this work is as follows: 
\subsubsection{Cyclic Matrix Transformation}
\label{subsubsec:16_cyclic}
In this technique, we transform each normalized feature vector into a $16\times16$ matrix. All the elements, $f_0$ to $f_{15}$ of a feature vector $F$ are placed as is in the first row, and each following row is a cyclic left shift of the row before it as depicted in Eq. \ref{eq_cyclic}. 
\begin{equation}
\label{eq_cyclic}
\begin{bmatrix}
f_0 & f_1 & f_2 & \cdots & f_{14} & f_{15} \\
f_1 & f_2 & f_3 & \cdots & f_{15} & f_0 \\
f_2 & f_3 & f_4 & \cdots & f_0 & f_1 \\
\vdots & \vdots & \vdots & \ddots & \vdots & \vdots \\
f_{14} & f_{15} & f_0 & \cdots & f_{12} & f_{13} \\
f_{15} & f_0 & f_1 & \cdots & f_{13} & f_{14}
\end{bmatrix}
\end{equation}
\subsubsection{Circulant Matrix Transformation}
\label{subsubsec:16_circ}
In a circulant matrix, each row has all of the elements constituting a vector, but a row/column is shifted one position to the right/down compared to the previous row/column. Circulant matrices can be described in different ways and have been proven useful in areas like numerical analysis and cryptography. These applications inspired us to use the circulant matrix as a transformation technique for our work. We create a $16\times16$ circulant matrix starting with a feature vector, $F$, which is placed in the first column of the matrix. Each following column is formed by shifting the elements of vector $F$ downwards, with the shift amount increasing by one for each new column. This process of turning a vector $F$ of $16$ elements into a circulant matrix is shown in Eq. \ref{eq_circulant}.
\begin{equation}
\label{eq_circulant}
\begin{bmatrix}
f_0 & f_{15} & f_{14} & \cdots & f_2 & f_1 \\
f_1 & f_0 & f_{15} & \cdots & f_3 & f_2 \\
f_2 & f_1 & f_1 & \cdots & f_4 & f_3 \\
\vdots & \vdots & \vdots & \ddots & \vdots & \vdots \\
f_{14} & f_{13} & f_{12} & \cdots & f_0 & f_{15} \\
f_{15} & f_{14} & f_{13} & \cdots & f_1 & f_0
\end{bmatrix}
\end{equation}

\begin{figure}
    \centering
    \begin{subfigure}[b]{0.2\textwidth}
        \centering
        \includegraphics[width=\textwidth]{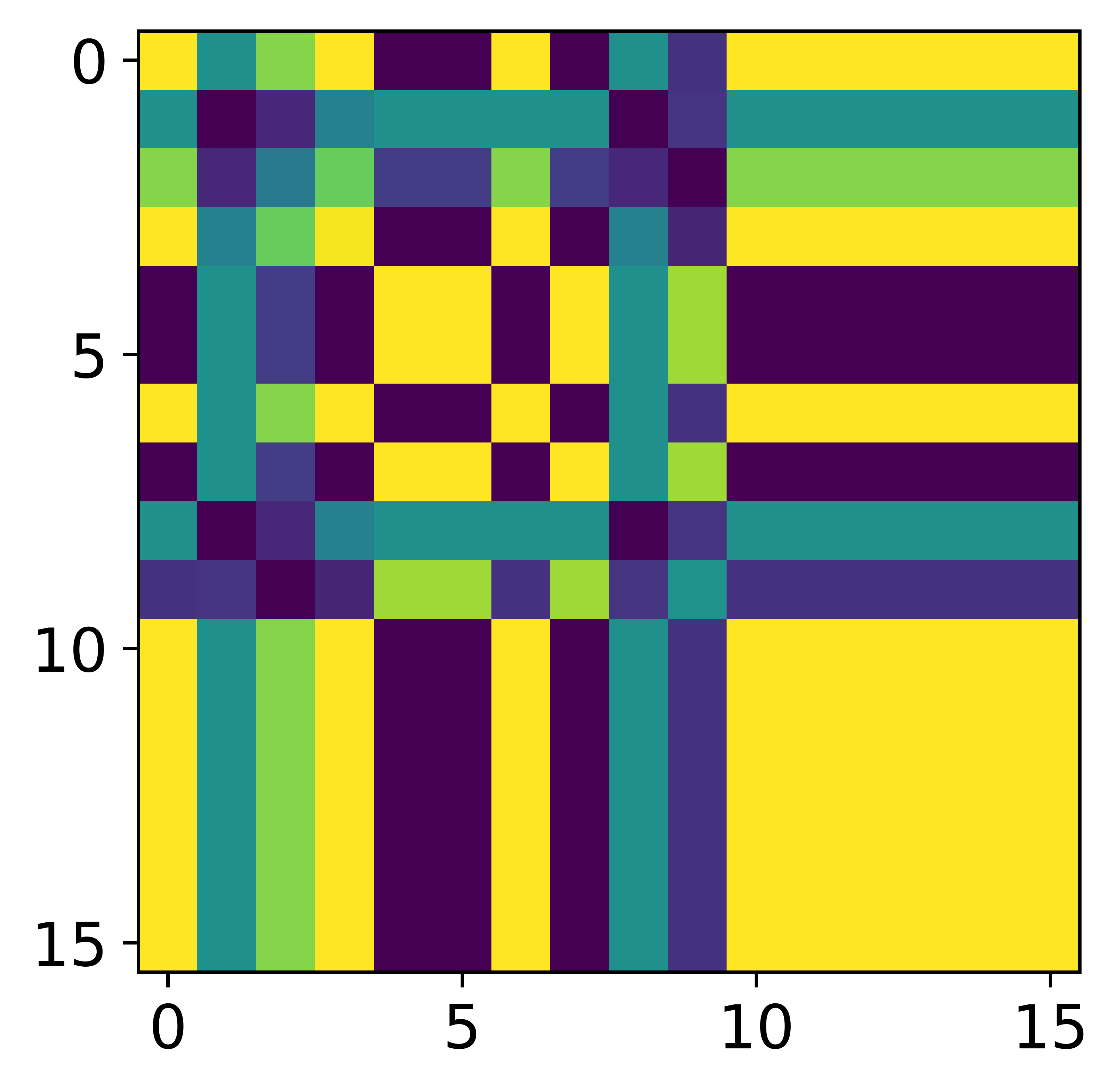}
        \caption{\textit{Normal Data}}
        \label{fig:gaf_image_normal}
    \end{subfigure}
    \begin{subfigure}[b]{0.2\textwidth}
        \centering
        \includegraphics[width=\textwidth]{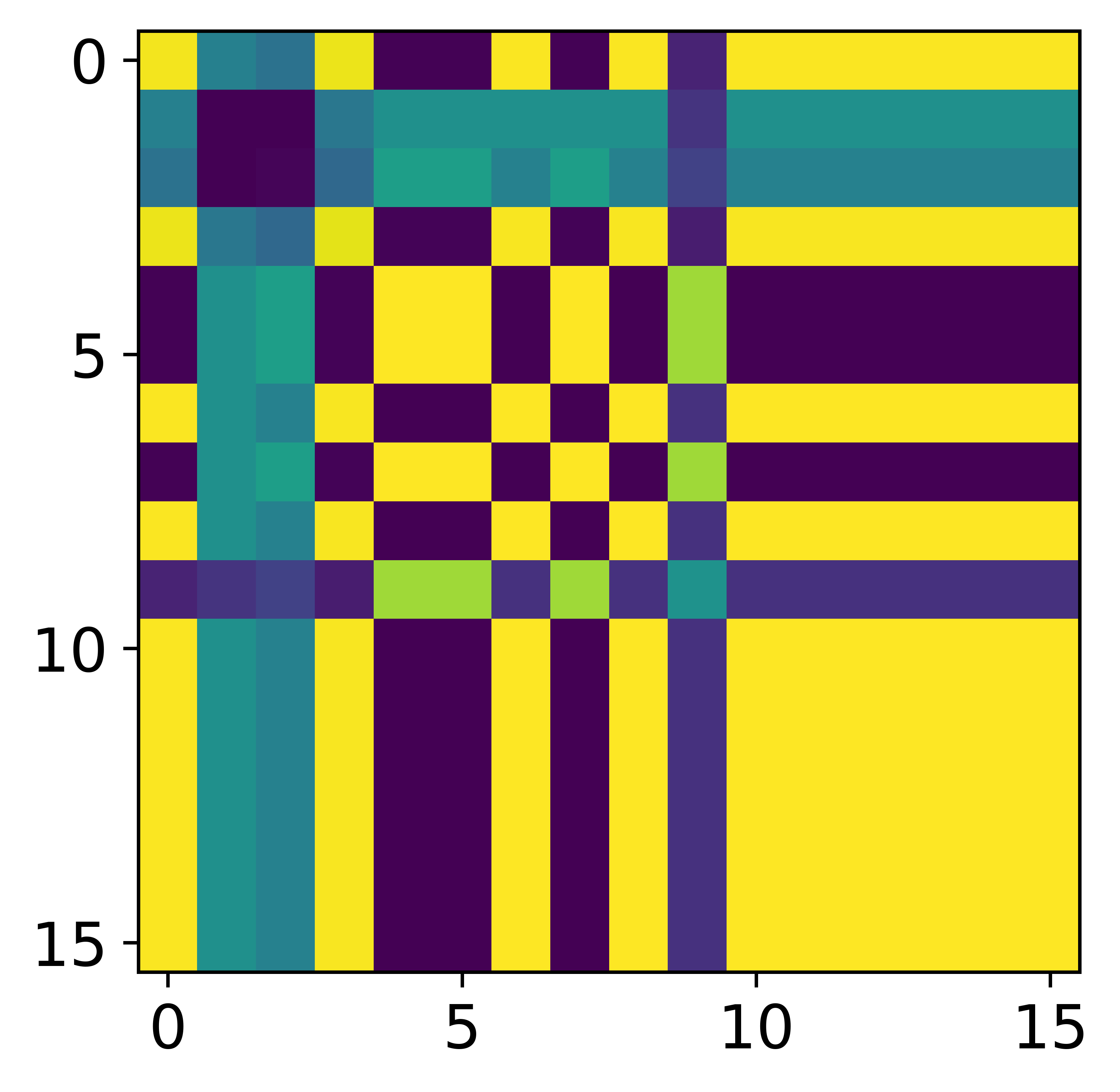}
        \caption{\textit{Deauthentication}}
        \label{fig:gaf_image_deauth}
    \end{subfigure}
    \begin{subfigure}[b]{0.2\textwidth}
        \centering
        \includegraphics[width=\textwidth]{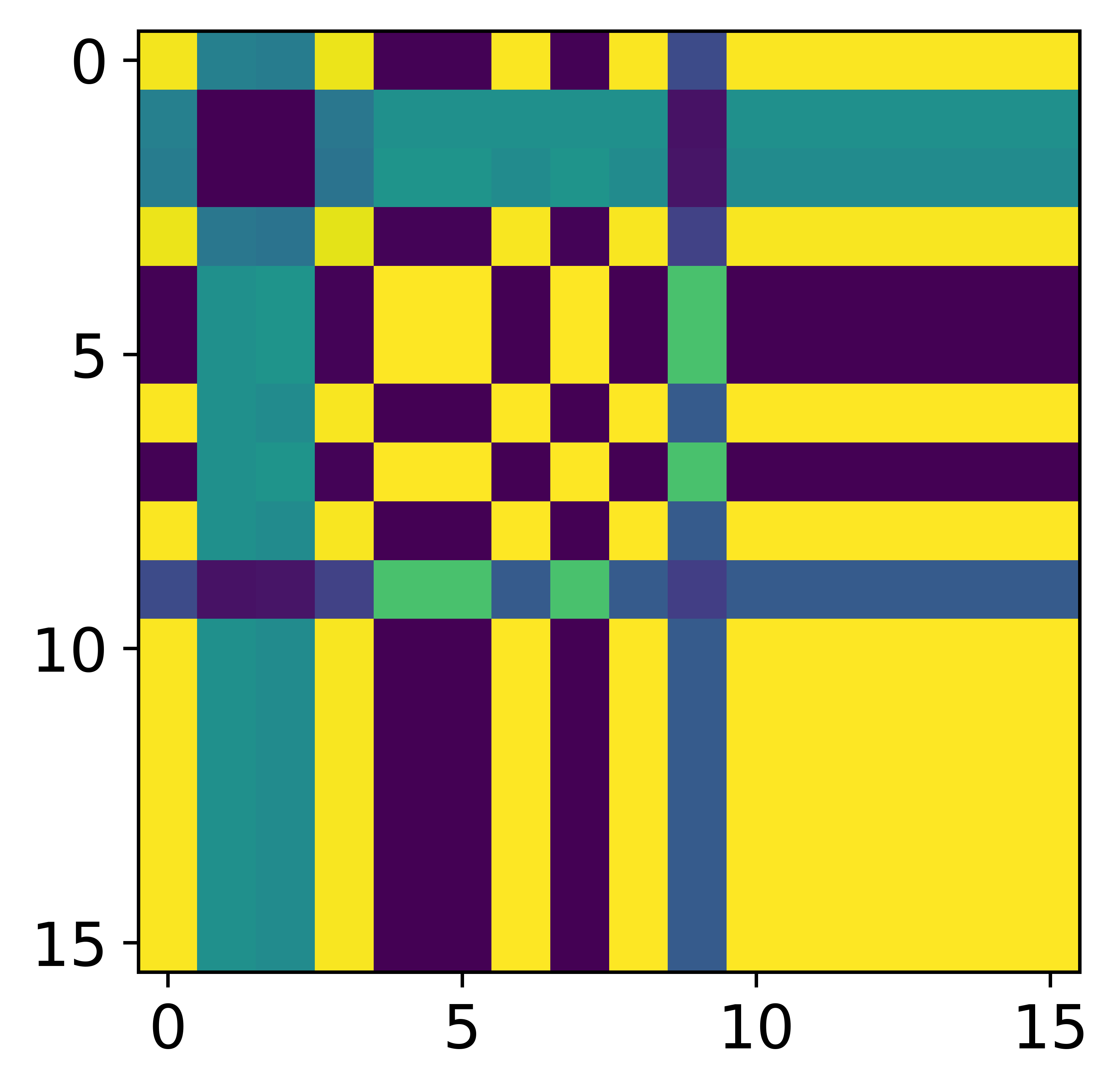}
        \caption{\textit{Disassociation}}
        \label{fig:gaf_image_disas}
    \end{subfigure}
    \begin{subfigure}[b]{0.2\textwidth}
        \centering
        \includegraphics[width=\textwidth]{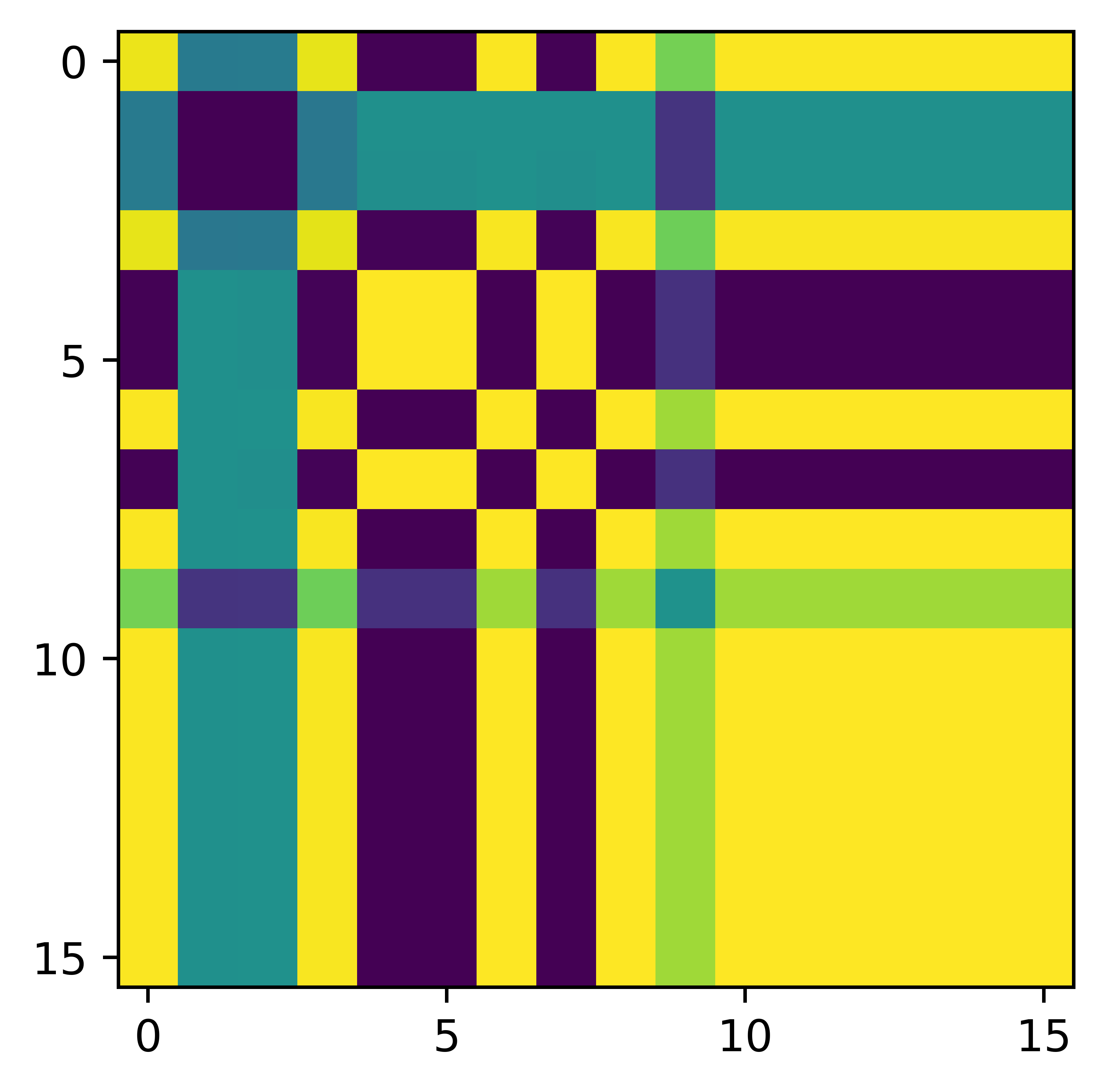}
        \caption{\textit{Re-Association}}
        \label{fig:gaf_image_reassoc}
    \end{subfigure}
    \begin{subfigure}[b]{0.2\textwidth}
        \centering
        \includegraphics[width=\textwidth]{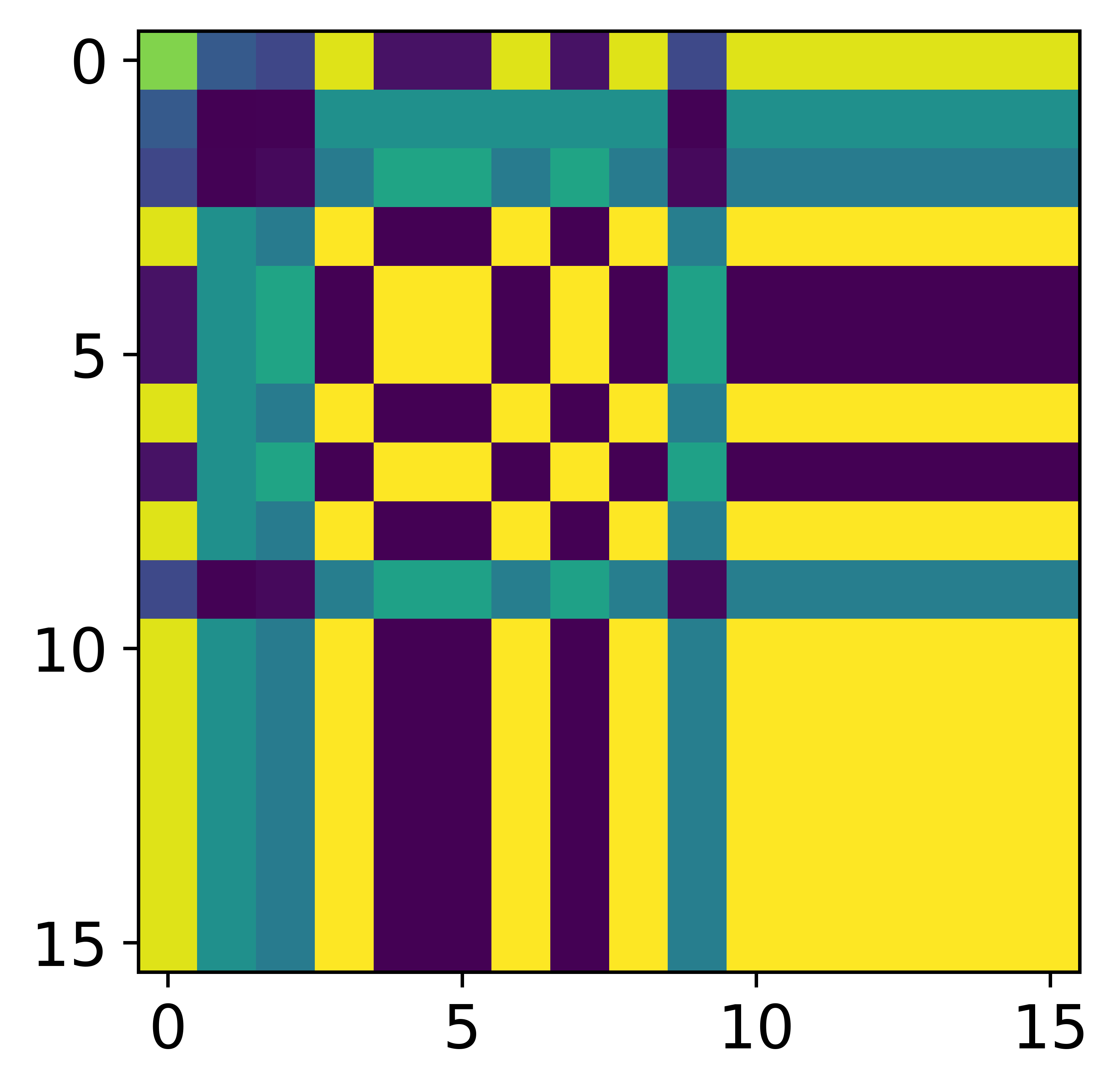}
        \caption{\textit{Rogue AP}}
        \label{fig:gaf_image_rogueap}
    \end{subfigure}
    \begin{subfigure}[b]{0.2\textwidth}
        \centering
        \includegraphics[width=\textwidth]{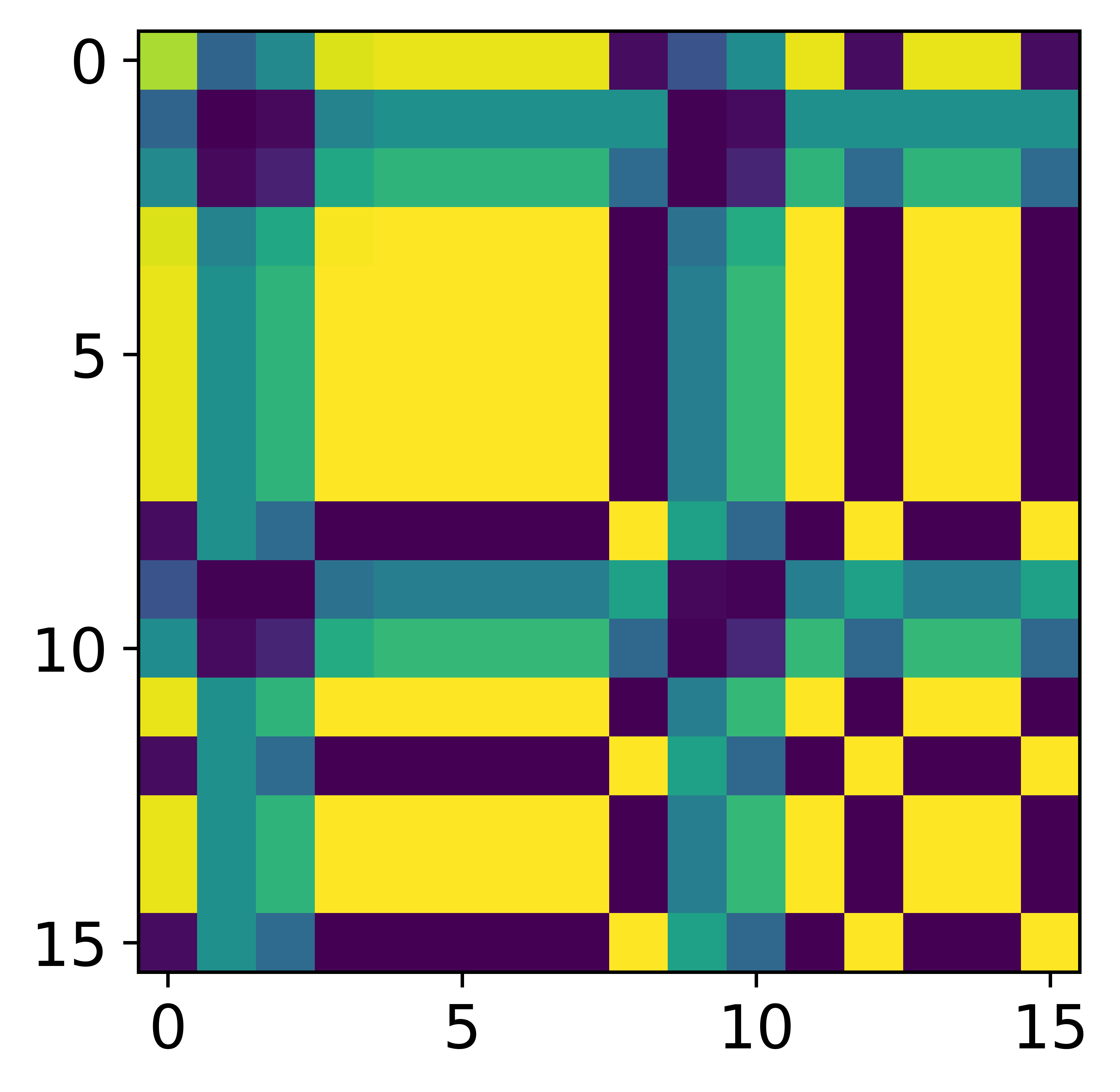}
        \caption{\textit{Krack}}
        \label{fig:gaf_image_krack}
    \end{subfigure}
    \begin{subfigure}[b]{0.2\textwidth}
        \centering
        \includegraphics[width=\textwidth]{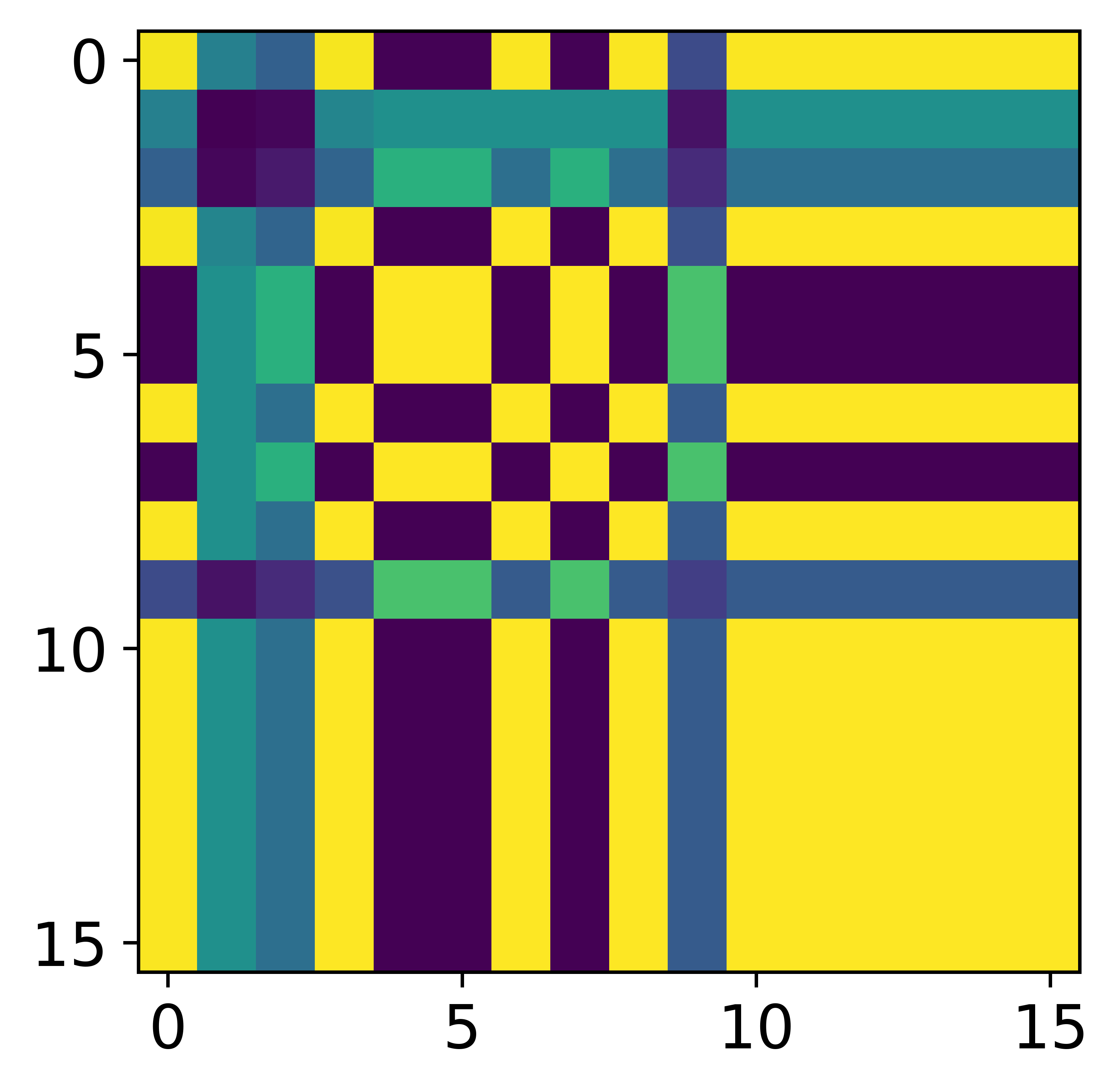}
        \caption{\textit{Kr00k}}
        \label{fig:gaf_image_kr00k}
    \end{subfigure}
    \begin{subfigure}[b]{0.2\textwidth}
        \centering
        \includegraphics[width=\textwidth]{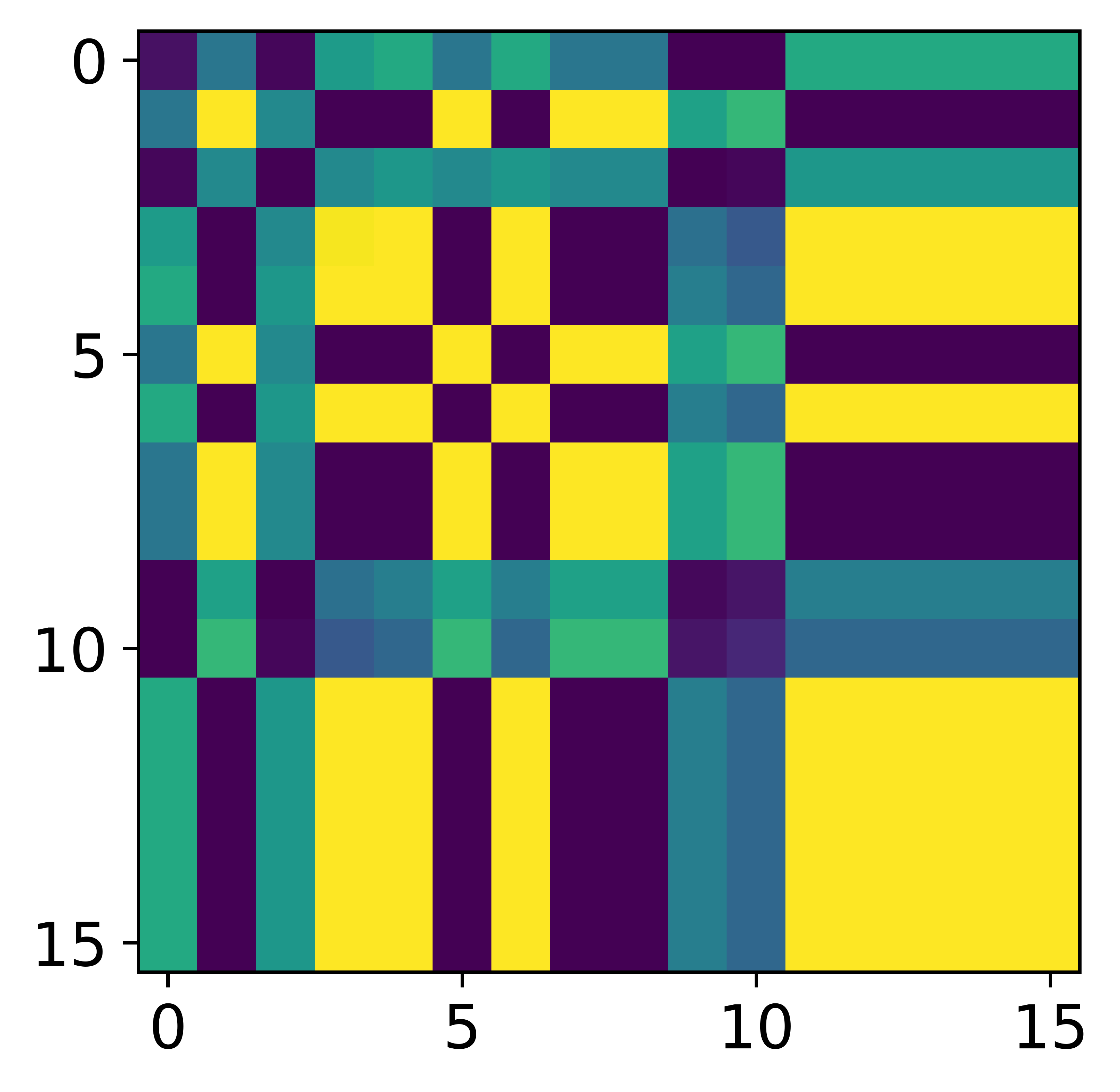}
        \caption{\textit{Evil Twin}}
        \label{fig:gaf_image_eviltwin}
    \end{subfigure}
    \caption{A few samples of GAF images for each type of attack data and normal data \cite{rayed}}
    \label{fig:gaf_images}
\end{figure}
\subsubsection{Grayscale Circulant Matrix Transformation}
\label{subsubsec:16_gray_circ}
In this technique, each feature in a traffic record is initially scaled to the range [0, 255] to emulate the pixel intensity range of a grayscale image. These normalized values are then embedded into a circulant matrix structure, after which the matrix entries are rescaled to fall within the range [-1, 1]. Although the procedure for generating the circulant matrix remains consistent, the intermediate scaling to [0, 255] modifies the original feature values, introducing variation in the final representation.
\subsubsection{Correlation Matrix Transformation}
\label{subsubsec:16_corr}
In this technique, we first create a $16\times16$ cyclic matrix using the feature vector of a traffic record similar to Eq. \ref{eq_cyclic}. Following that we compute the pairwise correlation for the columns $X$ and $Y$ from the cyclic matrix using Pearson Correlation Coefficient \cite{pearson_direct_2009}. For a given pair of columns ($X$, $Y$) in the matrix, we use Eq. \ref{eq_correlation} to compute the value of coefficient ($\rho$).
\begin{equation}
\label{eq_correlation}
    \rho_{X, Y} = \frac{\sum_{i=1}^{16}({X_i}-\overline{X})\times({Y_i}-\overline{Y})}{\sqrt{\sum_{i=1}^{16}({X_i}-\overline{X})} \times \sqrt{\sum_{i=1}^{16}({Y_i}-\overline{Y})}}
\end{equation}
 $X_i$ and $Y_i$ represent $i^{th}$ values in column X and Y; and $\overline{X}$ and $\overline{Y}$ represent the mean values for column X and Y. Finally, we create a $16 \times 16$ correlation matrix using the correlation coefficients as shown in Eq. \ref{eq_corr}.

 \begin{equation}
\label{eq_corr}
\begin{bmatrix}
\rho_{0,0} & \rho_{0,1} & \rho_{0,2} & \cdots & \rho_{0,14} & \rho_{0,15} \\
\rho_{1,0} & \rho_{1,1} & \rho_{1,2} & \cdots & \rho_{1,14} & \rho_{1,15} \\
\rho_{2,0} & \rho_{2,1} & \rho_{2,2} & \cdots & \rho_{2,14} & \rho_{2,15} \\
\vdots & \vdots & \vdots & \ddots & \vdots & \vdots \\
\rho_{14,0} & \rho_{14,1} & \rho_{14,2} & \cdots & \rho_{14,14} & \rho_{14,15} \\
\rho_{15,0} & \rho_{15,1} & \rho_{15,2} & \cdots & \rho_{15,14} & \rho_{15,15} \\
\end{bmatrix}
\end{equation}
\subsubsection{Gramian Angular Field (GAF) Matrix Transformation}
\label{subsubsec:gaf}
Several works have used GAF technique for features-to-image conversion for various signal processing and pattern recognition tasks \cite{gaf_encoding_2015, gaf_ecg_2019, gaf_eeg_2020, gaf_activitiy_recog_2020}. 
GAF employs the Gramian matrix to represent a signal or time-series data in polar coordinates as an image. GAF technique is commonly used with time-series data, however, in our case, we utilize the order of features in a traffic record as an indicator of data input received in a time-series. To transform a Wi-Fi traffic record into a GAF image, we first normalize the feature values within the range of [-1, 1]. These normalized values are then converted into the corresponding $cosine$ and $sine$ values in the polar coordinates, which are used to create the Gramian matrix. Each element of the Gramian matrix is calculated as the trigonometric sum or difference between a pair of features of the input in the polar coordinates. For our purpose, we utilize the summation variant to calculate the Gramian matrix, $G$ using Eq. 5 and 6: 
\begin{align}
    G & = \begin{bmatrix}
    cos\left( \phi_1 + \phi_1\right) & \cdots & cos\left( \phi_1 + \phi_n\right) \\
    cos\left( \phi_2 + \phi_1\right) & \cdots & cos\left( \phi_2 + \phi_n\right) \\
    \vdots & \ddots & \vdots \\
    cos\left( \phi_n + \phi_1\right) & \cdots & cos\left( \phi_n + \phi_n\right)
    \end{bmatrix} \\
    \cos{\left(\phi_i + \phi_j\right)} & = \cos{\left(\phi_i\right)} \cdot \cos{\left(\phi_j\right)} - \sin{\left(\phi_i\right)} \cdot \sin{\left(\phi_j\right)}
\end{align}
Fig. \ref{fig:gaf_images} displays \textit{sample} GAF images for each attack type and normal traffic in the AWID3 dataset. 
\subsection{Convolutional Neural Network (CNN) Model Architecture}
\label{subsec:model_architecture}
Since our work centers on implementing the IDS using a two-dimensional transformation of network features, we adopt a Convolutional Neural Network (CNN)—widely recognized as the most effective deep learning approach for processing image-like, two-dimensional data \cite{cnn_intro_2015}. Typically, a CNN architecture consists of alternating convolutional layers (employing non-linearities such as hyperbolic tangent) and subsampling (pooling) layers \cite{original_cnn}. In general, convolutional and pooling layers operate on the two-dimensional input data, however, one-dimensional CNN is not uncommon. Convolutional layers apply learnable filters (kernels) to the input matrix, capturing spatial patterns through the convolution operation. The number of filters used determines the number of feature maps (or channels) produced in the output. Each convolutional layer is typically followed by a non-linear activation function, similar to those used in traditional neural networks. Pooling layers, positioned between or after convolutional layers, reduce the spatial dimensions of the feature maps while preserving salient information. This is achieved through operations such as max pooling or average pooling, applied over local or global regions. The output size of the pooling operation depends on the filter size and the stride. After the convolutional and pooling stages, the resulting feature maps are flattened into a one-dimensional vector and passed through fully connected (dense) layers. These layers may contain multiple hidden layers, each with a configurable number of neurons, culminating in a final classification layer that produces the output. 

\begin{figure}
    \centering
    \includegraphics[width=0.75\linewidth]{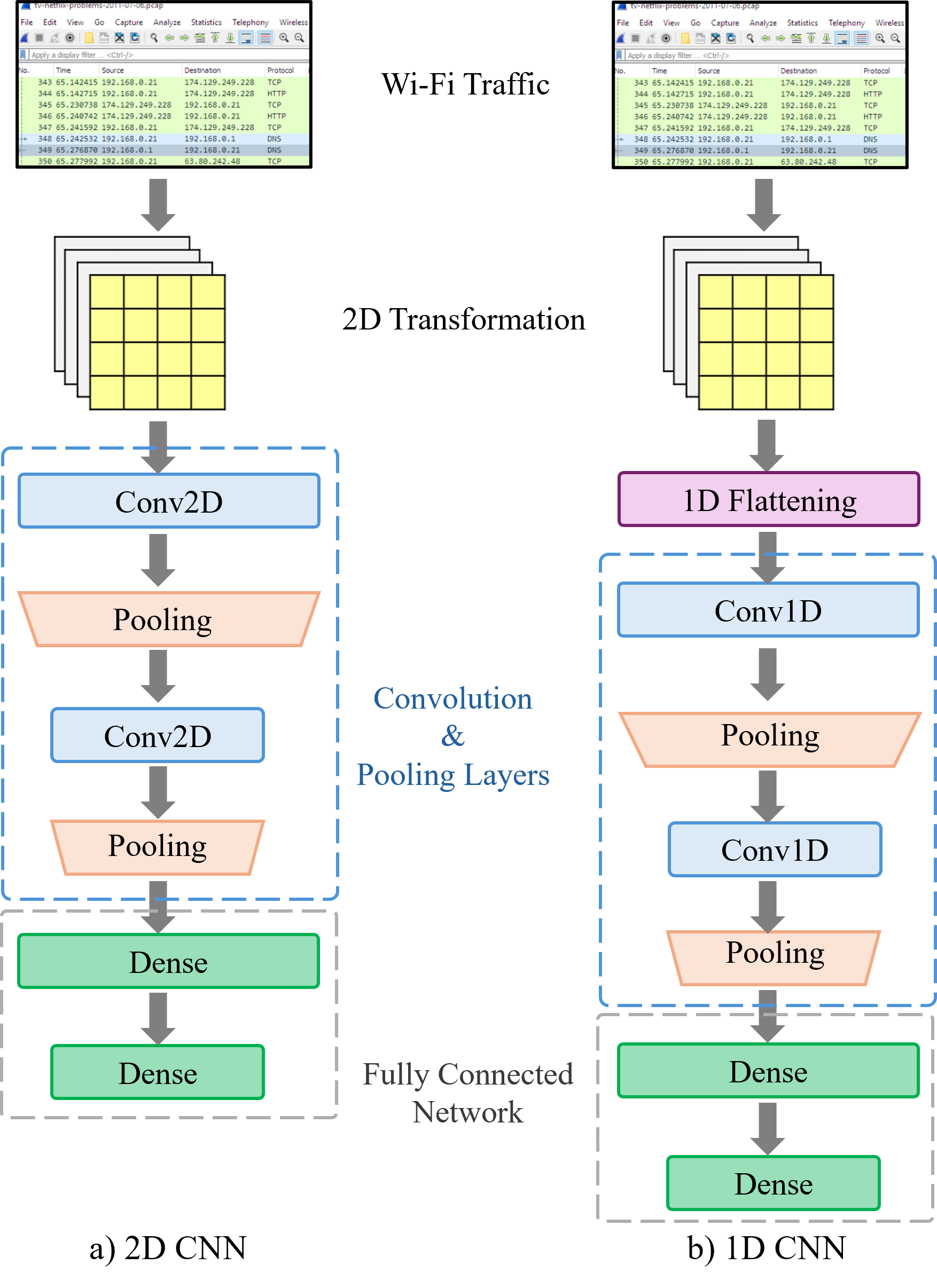}
    \caption{Architectures of 2D-CNN and 1D-CNN used in our work}
    \label{fig:cnn}
\end{figure}
The architecture of a CNN can be quite complex as evident in popular CNNs such as AlexNet \cite{alexnet_arch_2017}, and ResNet \cite{resnet_2015}. However, we implemented our IDS using four lightweight CNN models. Their architectures are discussed below. The convolution layers in each of them have 16 filter channels created by utilizing a 3x3 kernel matrix and ReLU activation function. Additionally, each model contains local averaging pooling layer(s) with a pooling matrix of size 2x2 and a stride of 2. 
\begin{itemize}
    \item \textbf{1D-CNN with 2 Convolution-Pooling layers pairs (1D-2L)}: This model takes the transformed matrix for each technique as input and flatten it before passing it on to the rest of the model. The model will have the following architecture: Convolution $\rightarrow$ Average Pooling $\rightarrow$ Convolution $\rightarrow$ Average Pooling. Fig. \ref{fig:cnn} (b) shows 1D-2L CNN architecture used in our work.
    \item \textbf{1D-CNN with 1 Convolution-Pooling layers pair (1D-1L)}: Similar to the previous model, this model takes the transformed matrix for each technique as input and flatten it before passing it on to the rest of the model. For this model we have a single Convolution layer followed by the Average Pooling layer as its architecture.
    \item \textbf{2D-CNN with 2 Convolution-Pooling layers pairs (2D-2L)}: This model takes the transformed matrix as input and passes it to the remaining CNN architecture laid out as follows: Convolution $\rightarrow$ Average Pooling $\rightarrow$ Convolution $\rightarrow$ Average Pooling. Fig. \ref{fig:cnn} (a) shows 2D-2L CNN architecture used in our work.
    \item \textbf{2D-CNN with 1 Convolution-Pooling layers pair (2D-1L)}: Similar to the previous model, this model takes the transformed matrix as is and passes it along to a Convolution layer followed by an Average Pooling layer.
\end{itemize}
Following the convolution and pooling layers, a 20\% dropout is applied, after which the feature maps are flattened and passed into a fully connected network comprising of a single hidden layer with 32 nodes, ReLU activation, and a dropout of 50\%. The dropout mechanism disables a subset of neurons in the preceding layer, preventing them from forwarding their outputs to the next layer. This regularization technique reduces reliance on specific neurons, thereby improving generalization and mitigating overfitting \cite{dropout_main_2012, dropout_theory_2013, dropout_cnn_2015}. After the fully connected layer, the classification layer executes the final prediction. Table \ref{tab:Hyperparameters} list the set of hyperparameters used to train the models. The maximum number of training epochs is set to 100, with early stopping enabled to halt training if validation performance does not improve over three consecutive epochs.
\begin{table}
\centering
\footnotesize
\caption{Hyperparameters used for training}
\label{tab:Hyperparameters}
%\resizebox{\columnwidth}{!}{
\begin{tabular}{p{4cm}|p{4cm}}
    \hline
    Parameter & Value \\ \hline \hline
    Optimizer & AMSGrad (variant of Adam \cite{amsgrad_adam_2019})
    \\ \hline
    Loss Metric & Binary Crossentropy (Binary) \newline Sparse Categorical Crossentropy  (Multiclass)
    \\ \hline
    Accuracy Metric & Binary Accuracy (Binary) \newline Sparse Categorical Accuracy (Multiclass)
    \\ \hline
    Hidden Layer and Convolution Layer(s) Activation Function & ReLU
    \\ \hline
    Classification Layer Nodes & 2 (Binary) \newline 8 (Multiclass)
    \\ \hline
    Classification Layer Activation Function & Sigmoid (Binary) \newline Softmax (Multiclass)
    \\ \hline
    Learning Rate & 0.001
    \\ \hline
    Epochs Trained & Variable (Max 100)
    \\ \hline
    Hidden Layer Nodes & 32
    \\ \hline
    Convolution Layer Filters & 16
    \\ \hline
    Convolution Layer Kernel Size & 3x3
    \\ \hline
    Pooling Layer Filter Size & 2x2
    \\ \hline
\end{tabular}
%}
\end{table}
\section{Results and Discussion}
\label{sec:results}
The final step of the ML pipeline was to train each model utilizing each one of the two-dimensional transformation techniques for binary and multi-class classifications. To identify the most effective model, we evaluate the performance of all trained models using the test dataset. The comparison is based on accuracy and F1-score. These two metrics collectively offer a comprehensive view of each model’s classification capability. These evaluation metrics are defined as follows and represented through Eq. 7--10:

\textbf{Accuracy}: It is the percentage of traffic records in the dataset, which are correctly classified by the model.
\begin{equation}
    \label{eq:accuracy}
    Accuracy\ (in\ \%)= \frac{TP+TN}{TP+TN+FP+FN} \times 100
\end{equation}
\textit{TP (True Positive)}: prediction of attack records in the dataset as attack records by the model, \textit{TN (True Negative)}: prediction of normal records as normal records, \textit{FP (False Positive)}: prediction of normal records as attack records, and \textit{FN (False Negative)}: prediction of attack records as normal records.

\textbf{F1-score}: It represents the harmonic mean of \textit{Precision} and \textit{Recall}, assigning equal importance to both metrics in evaluating a model’s performance. 
\begin{align}
Precision\ (in\ \%) & = \frac{TP}{TP+FP} \times 100\\
Recall\ (in\ \%) & = \frac{TP}{TP+FN} \times 100\\ 
F1-score\ (in\ \%) & = \frac{2 \times Precision \times Recall}{Precision+Recall} \times 100
\end{align}

\textit{Precision} measures the percentage of records predicted as attacks by the model that are indeed actual attack records in the dataset, while \textit{Recall} quantifies the percentage of true attack records in the dataset that are correctly identified by the model.
\begin{table}[!hbtp]
\begin{center}
\footnotesize
\centering
\caption{Performance metrics for binary CNN models}
\setlength{\tabcolsep}{3pt}
\begin{tabular}{p{50pt} p{30pt} p{30pt} p{30pt} p{31pt} p{30pt} p{40pt}}
\hline 
 &  & \multicolumn{2}{c}{\textbf{Accuracy}} &  \\
\cline{3-4}
\textbf{Technique} & \textbf{Model} & \textbf{Train} & \textbf{Test} & \textbf{Precision} & \textbf{Recall} & \textbf{F1-score} \\ \hline \hline
\multirow{4}{*}{{\centering\textbf{Cyclic}}}
& 1D-2L
& 99.74\% & 99.80\% & 99.55\% & 98.95\% & 99.25\% \\
\cline{2-7}
& 1D-1L
& 99.62\% & 99.59\% & 97.11\% & 99.89\% & 98.48\% \\
\cline{2-7}
& \textit{2D-2L}
& \textit{99.75\%} & \textit{99.80\%} & \textit{99.39\%} & \textit{99.12\%} & \textit{99.25\%} \\
\cline{2-7}
& 2D-1L
& 99.57\% & 99.63\% & 99.09\% & 98.15\% & 98.62\% \\ \hline \hline

\multirow{4}{*}{{\centering\textbf{Circulant}}}
& 1D-2L
& 99.71\% & 99.73\% & 98.96\% & 99.00\% & 98.98\% \\
\cline{2-7}
& 1D-1L
& 99.54\% & 99.55\% & 97.21\% & 99.47\% & 98.33\% \\
\cline{2-7}
& \textit{2D-2L}
& \textit{99.76\%} & \textit{99.80\%} & \textit{99.62\%} & \textit{98.87\%} & \textit{99.25\%} \\
\cline{2-7}
& 2D-1L
& 99.60\% & 99.59\% & 97.24\% & 99.75\% & 98.48\% \\ \hline \hline

\multirow{4}{*}{\parbox{40pt}{\centering\textbf{Grayscale}\\ \textbf{Circulant}}}
& \textit{1D-2L}
& \textit{99.76\%} & \textit{99.82\%} & \textit{99.21\%} & \textit{99.41\%} & \textit{99.31\%} \\
\cline{2-7}
& 1D-1L
& 99.51\% & 99.60\% & 97.17\% & 99.90\% & 98.51\% \\
\cline{2-7}
& 2D-2L
& 99.72\% & 99.80\% & 99.60\% & 98.93\% & 99.27\% \\
\cline{2-7}
& 2D-1L
& 99.56\% & 99.60\% & 97.21\% & 99.89\% & 98.53\% \\ \hline \hline

\multirow{4}{*}{{\centering\textbf{Correlation}}}
& \textit{1D-2L}
& \textit{99.53\%} & \textit{99.67\%} & \textit{98.05\%} & \textit{99.53\%} & \textit{98.79\%} \\
\cline{2-7}
& 1D-1L
& 98.88\% & 99.26\% & 96.03\% & 98.54\% & 97.27\% \\
\cline{2-7}
& 2D-2L
& 99.59\% & 99.57\% & 99.57\% & 97.23\% & 98.38\% \\
\cline{2-7}
& 2D-1L
& 98.98\% & 99.23\% & 95.41\% & 99.00\% & 97.17\% \\ \hline \hline

\multirow{4}{*}{{\centering\textbf{GAF}}}
& 1D-2L
& 99.90\% & 99.92\% & 99.70\% & 99.67\% & 99.69\% \\
\cline{2-7}
& 1D-1L
& 99.67\% & 99.83\% & 98.95\% & 99.81\% & 99.39\% \\
\cline{2-7}
& \textbf{2D-2L}
& \textbf{99.86\%} & \textbf{99.93\%} & \textbf{99.90\%} & \textbf{99.56\%} & \textbf{99.73\%} \\
\cline{2-7}
& 2D-1L
& 99.50\% & 99.58\% & 97.18\% & 99.77\% & 98.46\% \\ \hline \hline

\end{tabular}
\label{tab:perf_binary}
\end{center}
\end{table}
\begin{table}[!hbtp]
\begin{center}
\footnotesize
\centering
\caption{Performance metrics for multi-class CNN models}
\setlength{\tabcolsep}{3pt}
\begin{tabular}{p{50pt} p{30pt} p{30pt} p{30pt} p{31pt} p{30pt} p{40pt}}
\hline 
 &  & \multicolumn{2}{c}{\textbf{Accuracy}} &  \\
\cline{3-4}
\textbf{Technique} & \textbf{Model} & \textbf{Train} & \textbf{Test} & \textbf{Precision} & \textbf{Recall} & \textbf{F1-score} \\ \hline \hline

\multirow{4}{*}{{\centering\textbf{Cyclic}}}
& \textit{1D-2L}
& \textit{99.14\%} & \textit{99.49\%} & \textit{99.50\%} & \textit{99.49\%} & \textit{99.49\%} \\
\cline{2-7}
& 1D-1L
& 97.16\% & 98.30\% & 98.44\% & 98.30\% & 98.02\% \\
\cline{2-7}
& 2D-2L
& 98.99\% & 99.33\% & 99.33\% & 99.33\% & 99.31\% \\
\cline{2-7}
& 2D-1L
& 97.72\% & 97.92\% & 98.10\% & 97.92\% & 97.58\% \\ \hline \hline

\multirow{4}{*}{{\centering\textbf{Circulant}}}
& \textit{1D-2L}
& \textit{99.24\%} & \textit{99.48\%} & \textit{99.48\%} & \textit{99.48\%} & \textit{99.47\%} \\
\cline{2-7}
& 1D-1L
& 97.89\% & 98.76\% & 98.88\% & 98.76\% & 98.69\% \\
\cline{2-7}
& 2D-2L
& 99.12\% & 99.41\% & 99.24\% & 99.41\% & 99.40\% \\
\cline{2-7}
& 2D-1L
& 98.33\% & 99.10\% & 99.15\% & 99.10\% & 99.07\% \\ \hline \hline

\multirow{4}{*}{\parbox{40pt}{\centering\textbf{Grayscale}\\ \textbf{Circulant}}}
& 1D-2L
& 98.50\% & 99.31\% & 99.32\% & 99.31\% & 99.30\% \\
\cline{2-7}
& 1D-1L
& 97.62\% & 99.17\% & 99.16\% & 99.17\% & 99.13\% \\
\cline{2-7}
& \textit{2D-2L}
& \textit{99.08\%} & \textit{99.40\%} & \textit{99.41\%} & \textit{99.40\%} & \textit{99.39\%} \\
\cline{2-7}
& 2D-1L
& 98.05\% & 99.00\% & 99.12\% & 99.00\% & 99.00\% \\ \hline \hline

\multirow{4}{*}{{\centering\textbf{Correlation}}}
& 1D-2L
& 98.47\% & 99.11\% & 99.19\% & 99.11\% & 99.12\% \\
\cline{2-7}
& 1D-1L
& 96.87\% & 99.26\% & 97.99\% & 97.69\% & 97.45\% \\
\cline{2-7}
& \textit{2D-2L}
& \textit{98.97\%} & \textit{99.34\%} & \textit{99.35\%} & \textit{99.34\%} & \textit{99.33\%} \\
\cline{2-7}
& 2D-1L
& 96.70\% & 97.60\% & 97.78\% & 97.60\% & 97.24\% \\ \hline \hline

\multirow{4}{*}{{\centering\textbf{GAF}}}
& \textbf{1D-2L}
& \textbf{99.36\%} & \textbf{99.62\%} & \textbf{99.62\%} & \textbf{99.62\%} & \textbf{99.62\%} \\
\cline{2-7}
& 1D-1L
& 97.81\% & 99.05\% & 99.13\% & 99.05\% & 99.03\% \\
\cline{2-7}
& 2D-2L
& 99.28\% & 99.50\% & 99.51\% & 99.50\% & 99.48\% \\
\cline{2-7}
& 2D-1L
& 98.41\% & 99.32\% & 99.30\% & 99.32\% & 99.27\% \\ \hline \hline

\end{tabular}
\label{tab:perf_multi}
\end{center}
\end{table}
Once a model is trained for binary or multi-class classification, it is evaluated on 1,046,692 records from the test dataset. Performance metrics are recorded along with the total prediction time for all samples. Table \ref{tab:perf_binary} presents the accuracy and F1-score for each model trained for binary classification, while Table \ref{tab:perf_multi} reports the corresponding metrics for multi-class classification.

\begin{figure}[!t]
    \centering
    \begin{subfigure}[b]{0.6\textwidth}
        \centering
        \includegraphics[width=\textwidth]{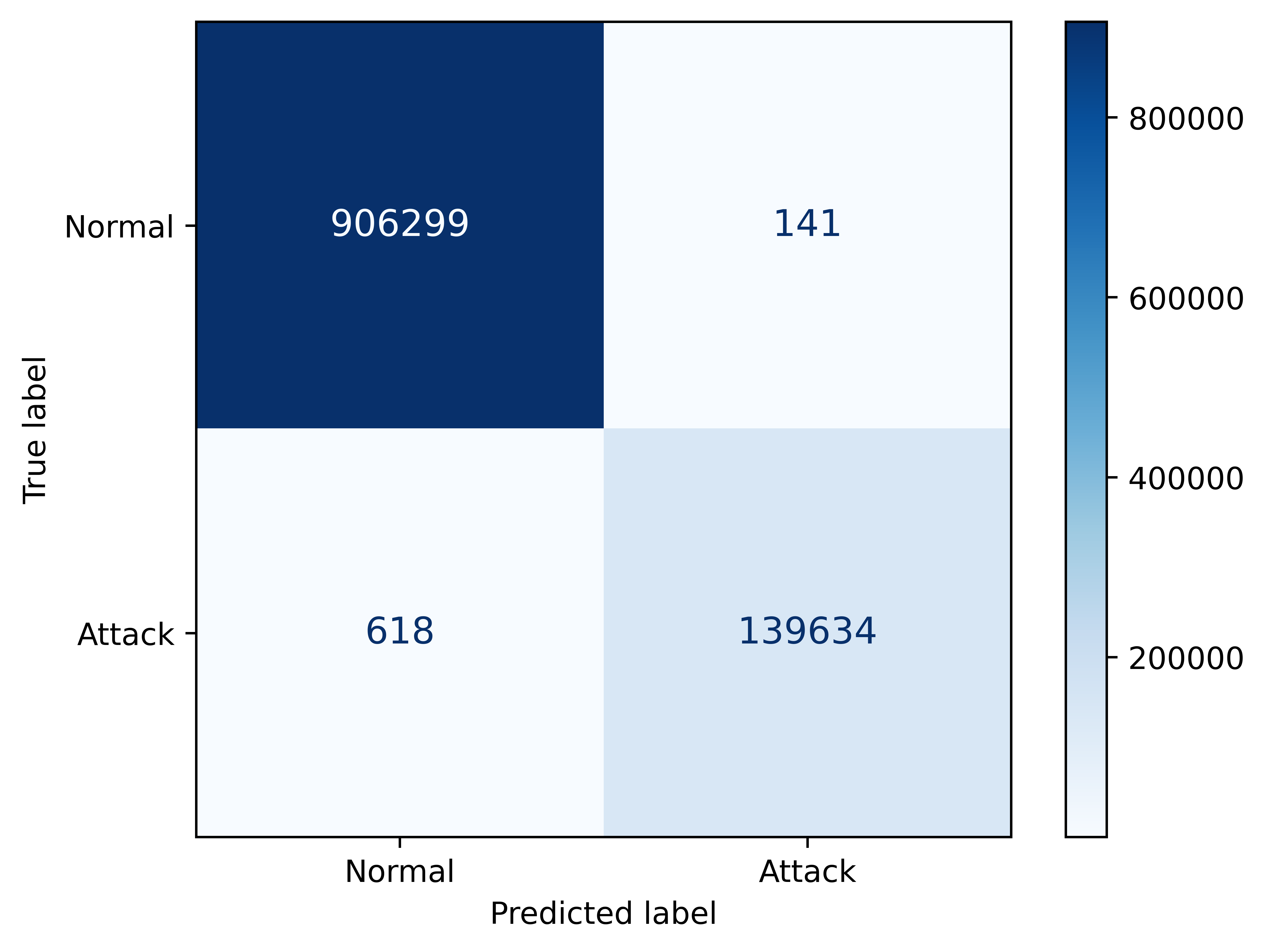}
        \caption{\textit{GAF based 2D-2L CNN model for binary classification}}
        \label{fig:cm_binary_best}
    \end{subfigure}
    \begin{subfigure}[b]{0.75\textwidth}
        \centering
        \includegraphics[width=\textwidth]{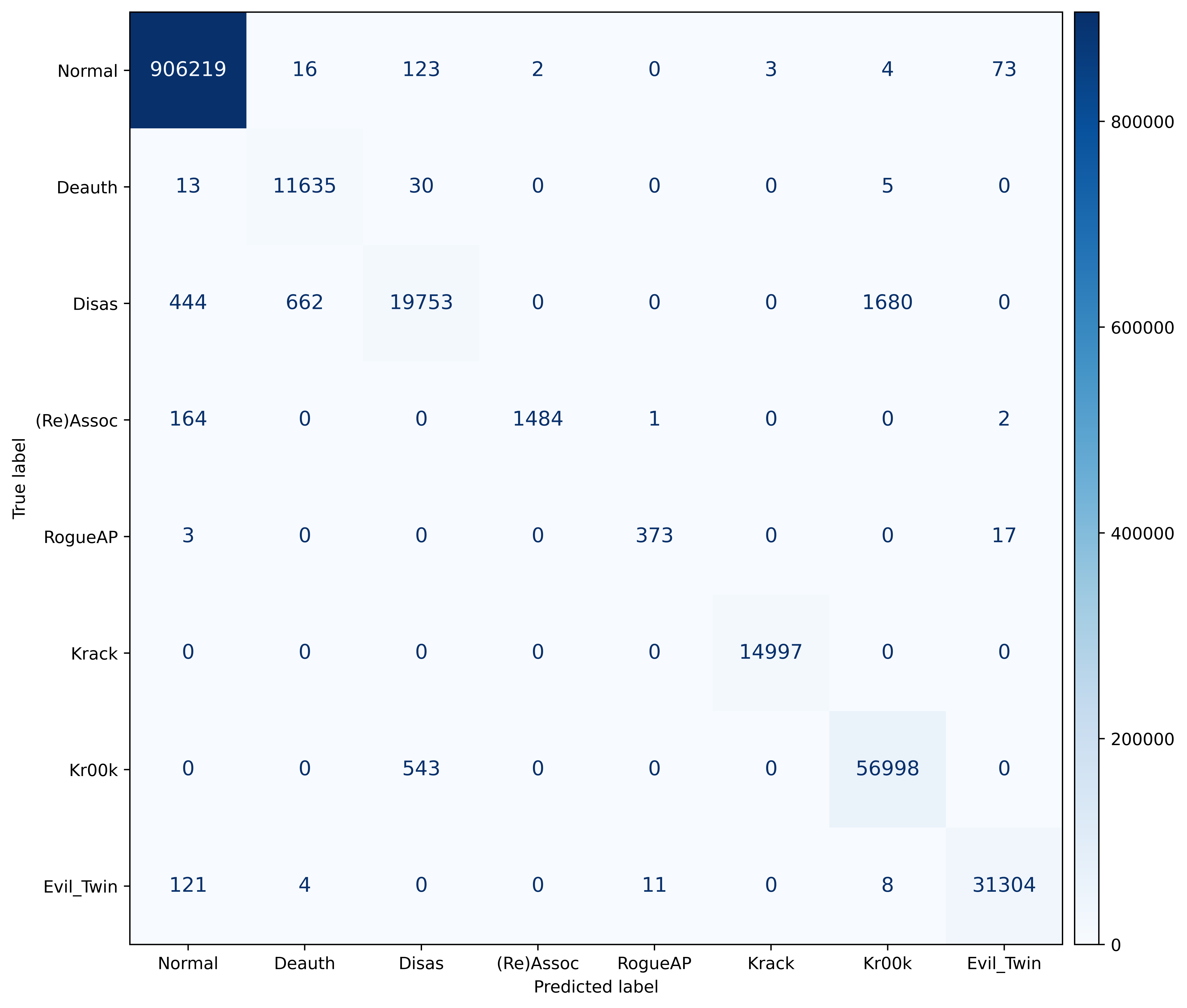}
        \caption{\textit{GAF based 1D-2L CNN model for multi-class classification}}
        \label{fig:cm_multi_best}
    \end{subfigure}
    \caption{Confusion matrices for binary and multi-class classifications for best performing transformation techniques and CNN models}
    \label{fig:cm_binary_best_chosen}
\end{figure}
For binary classification of attack and normal traffic records, it can be observed from Table~\ref{tab:perf_binary} that the GAF-based transformation technique combined with the 2D-2L CNN model achieved the highest performance in both training and testing phases. The model attained an average training accuracy of 99.86\% and a testing accuracy of 99.93\%. The corresponding F1-score was 99.73\%, with precision and recall values of 99.90\% and 99.56\%, respectively. Among all CNN models evaluated, the GAF technique consistently yielded the best results, with an average testing accuracy of 99.81\% and an F1-score of 99.32\%. Figure \ref{fig:cm_binary_best} shows the confusion matrix for GAF based 2D-2L CNN model for binary classification.

Table \ref{tab:perf_multi} presents the multi-class classification performance of various CNN models across different matrix transformation techniques. The GAF-based 1D-2L CNN model achieved the highest performance, with training and testing accuracies of 99.36\% and 99.62\%, respectively. The corresponding F1-score on the test dataset was 99.62\%. Among the four CNN models evaluated, the GAF approach consistently outperformed the others, yielding an average testing accuracy of 99.37\% and an F1-score of 99.35\%. The confusion matrix for the GAF-based 1D-2L CNN model, shown in Figure \ref{fig:cm_multi_best}, further illustrates its strong multi-class classification capability.

CNN models typically employed in computer vision and image processing tasks are often deep and computationally intensive, incorporating numerous convolution and pooling layers. In this work, lightweight CNN architectures were adopted to minimize training and inference times, as well as model size, making them suitable for real-time intrusion detection in Wi-Fi networks. Model training was conducted on a workstation equipped with an AMD Ryzen 9 5900X 12-Core Processor (3.7 GHz, 12 Cores, 24 Threads), 32 GB RAM, and an NVIDIA GeForce RTX 3070 GPU. The training duration for the 2D-2L CNN model (binary classification) and the 1D-2L CNN model (multi-class classification) was $\approx$ 250 minutes. Given the system specifications, these times are not computationally prohibitive for DL model training. For inference on 1,046,692 test records, the 2D-2L model required $\approx$ 50,400 ms (i.e., $\approx$ 48 $\mu$s per record), whereas the 1D-2L model completed prediction in $\approx$ 42,400 ms (i.e., $\approx$ 40 $\mu$s per record). The total number of trainable parameters in the 2D-2L and 1D-2L models were approximately 11,000 and 34,000, respectively—significantly lower than conventional CNN architectures such as AlexNet (62.4M parameters) and ResNet (25.6M parameters). 

Table \ref{tab:related_comparison} compares our work with selected existing studies. In binary classification, the GAF based 2D-2L model achieved the highest accuracy and F1-score among the referenced works. For multi-class classification, traffic was categorized into classes ranging from three to 19. Although a direct comparison is not feasible due to differing class definitions, the GAF based 1D-2L CNN model attained the highest F1-score of 99.62\%. In terms of accuracy, the best results were reported in \cite{awid_features_2022} and \cite{bhutta2024lightweight}, both achieving 99.96\%. However, the F1-score for 1D-2L was the highest and more reliable for imbalanced dataset. The source code of our implementation is avaliable as open source in \cite{rayed_dl_wids}. 
\begin{table}
\centering
\footnotesize
\caption{Comparison of our work with a few existing ones on AWID3 dataset}
\setlength{\tabcolsep}{2pt}
\begin{tabular}{p{2.75cm} p{2cm} p{1.75cm} p{1.75cm}}
    \hline
    \textbf{Model} & \textbf{Classification} & \textbf{Accuracy (\%)} & \textbf{F1-score (\%)} \\ \hline \hline

    HT \cite{multiple_ml_awid3_2021} & Binary & 98.00 & -- \\ \hline
    CNN \cite{yonbawi2025transferability} & Binary & 97.28 & 98.53 \\ \hline
    \textbf{2D-2L with GAF} & Binary & \textbf{99.93} & \textbf{99.73} \\ \hline \hline
    ET \cite{awid_features_2022} & Multi-class (3) & \textbf{99.96} & 99.52
    \\ \hline
    Bagging \cite{ensemble_ml_awid3_2022} & Multi-class (3) & 98.73 & 88.07 \\ \hline 
    LightGBM \cite{kampourakis2025balancing} & Multi-class (3) & 99.96 & 98.95 \\ \hline 
    LightGBM \cite{bhutta2024lightweight} & Multi-class (19) & 99.77 & 94.79 \\ \hline
    DT \cite{bhutta2025advancing} & Multi-class (19) & 99.45 & 97.00 \\ \hline
    \textbf{1D-2L with GAF}  & Multi-class (8) & 99.62 & \textbf{99.62} 
    \\ \hline
\end{tabular}
\label{tab:related_comparison}
\end{table}
\section{Conclusion}
\label{sec:conclusion}
In this work, we implemented CNN-based DL models for intrusion detection in Wi-Fi networks using the benchmark AWID3 dataset. A selected subset of link layer headers extracted from Wi-Fi frame data was utilized as input features and reshaped into a 16$\times$16 single-channel matrix for the DL models. We applied five distinct transformation techniques and evaluated four lightweight CNN architectures. Among these, the 2D-2L CNN model combined with the GAF transformation achieved an accuracy of 99.93\% and an F1-score of 99.73\% for binary classification of traffic records. For multi-class classification involving specific intrusions and normal traffic, the 1D-2L CNN model attained an accuracy of 99.62\% and an F1-score of 99.62\%. The best-performing models demonstrated inference times of approximately 48$\mu$s and 40$\mu$s per traffic record, indicating their suitability for real-time intrusion detection. In future work, we plan to deploy these models on programmable network devices (e.g., OpenWrt).  
\bibliographystyle{IEEEtran}
\bibliography{references}
\end{document}